\pdfoutput=1
\documentclass{article}
\usepackage{microtype}
\usepackage{graphicx}
\usepackage{subfigure}
\usepackage{booktabs} 
\usepackage{amssymb}
\usepackage{array}
\usepackage{amsmath}
\usepackage{bm}
\usepackage{bbm}
\usepackage{algorithm}
\usepackage{algorithmic}
\usepackage{caption}
\usepackage[abs]{overpic}
\usepackage[dvipsnames]{xcolor}

\newcounter{nbdrafts}
\makeatletter
\newcommand{\checknbdrafts}{
\ifnum \thenbdrafts > 0
\@latex@warning@no@line{**********************************************************************}
\@latex@warning@no@line{* The document contains \thenbdrafts \space draft note(s)}
\@latex@warning@no@line{**********************************************************************}
\fi}

\makeatother

\newcommand{\bA}[0]{\mathbf{A}}
\newcommand{\bC}[0]{\mathbf{C}}
\newcommand{\bX}[0]{\mathbf{X}}

\newcommand{\bY}[0]{\mathbf{Y}}

\newcommand{\bM}[0]{\mathbf{M}}
\newcommand{\bP}[0]{\mathbf{P}}

\newcommand{\bR}[0]{\mathbb{R}}
\newcommand{\bE}[0]{\mathbb{E}}

\newcommand{\mC}[0]{\mathcal{C}}
\newcommand{\mE}[0]{\mathcal{E}}

\newcommand{\mF}[0]{\mathcal{F}}

\newcommand{\mM}[0]{\mathcal{M}}

\newcommand{\mN}[0]{\mathcal{N}}

\newcommand{\mG}[0]{\mathcal{G}}
\newcommand{\mR}[0]{\mathcal{R}}

\newif\ifdraft
\drafttrue

\ifdraft
 \newcommand{\PF}[1]{\textcolor{blue}{{\bf PF: #1}}}
 \newcommand{\pf}[1]{\textcolor{blue}{#1}}
 \newcommand{\PB}[1]{\textcolor{red}{{\bf PB: #1}}}

 \newcommand{\FF}[1]{\textcolor{red}{{\bf FF: #1}}}

\else
 \newcommand{\PF}[1]{}
 \newcommand{\FF}[1]{}
 \newcommand{\PB}[1]{}
 \newcommand{\pf}[1]{ #1 }

\fi

\DeclareMathOperator*{\argmin}{\arg\!\min}

\newcommand\blfootnote[1]{%
  \begingroup
  \renewcommand\thefootnote{}\footnote{#1}%
  \addtocounter{footnote}{-1}%
  \endgroup
}

\usepackage[accepted]{icml2018}

\icmltitlerunning{Geodesic Convolutional Shape Optimization}

\begin{document}

\twocolumn[
\icmltitle{Geodesic Convolutional Shape Optimization}



\icmlsetsymbol{equal}{*}

\begin{icmlauthorlist}
\icmlauthor{Pierre Baque}{equal,epfl}
\icmlauthor{Edoardo Remelli}{equal,epfl}
\icmlauthor{Fran\c cois Fleuret}{idiap,epfl}
\icmlauthor{Pascal Fua}{epfl}
\end{icmlauthorlist}

\icmlaffiliation{epfl}{CVLab, EPFL, Lausanne, Switzerland}

\icmlaffiliation{idiap}{Machine Learning Group, Idiap, Martigny, Switzerland}

\icmlcorrespondingauthor{Pierre Baque}{pierre.baque@epfl.ch}

\icmlkeywords{Geometric Convolutional Networks, Shape optimization}

\vskip 0.3in
]



\printAffiliationsAndNotice{\icmlEqualContribution} 

\begin{abstract}

Aerodynamic shape optimization has many industrial applications. Existing methods, however, are so computationally demanding that typical engineering practices are to either simply try a limited number of hand-designed shapes or restrict oneself to shapes that can be parameterized using only few degrees of freedom.  

In this work, we introduce a new way to optimize complex shapes fast and accurately. To this end, we train Geodesic Convolutional Neural Networks to emulate a fluidynamics simulator. The key to making this approach practical is remeshing the original shape using a poly-cube map, which makes it possible to perform the computations on GPUs instead of CPUs.  The neural net is then used to formulate an objective function that is differentiable with respect to the shape parameters, which can then be optimized using a gradient-based technique.  This outperforms state-of-the-art methods by 5 to 20\% for standard problems and, even more importantly, our approach applies to cases that previous methods cannot handle.

\end{abstract}

\comment{

In this work we try to alleviate the problems cited above. We introduce a new type of deep convolutional neural network able to compute convolutions efficiently over a geodesic surface described by a mesh. We train such a Geodesic Convolutional Neural Network (GCNN) to predict the aerodynamic performance measure of interest (e.g. CX, drag, lift), and then use this GCNN predictor as a differentiable black-box surrogate model mapping a shape design to an objective. 

We optimize the shape given as input to this surrogate with gradient descent, resulting in a new optimisation method, non-parametric, and able to take into account additional shape constraints in a very flexible way. Experimental validation, on 2D and 3D aerodynamic optimization problems, demonstrate that it makes optimization possible for large parameters spaces, where previous approaches fail. 

	Aerodynamic shape optimization is an unsolved problem with vast industrial application. Over the years, several classes of methods were developed, and some of the most successful ones are based on Meta-Models which interpolate the computations made by a fluid simulator. However, such methods have the inconvenient that they assume a low dimensional parametrization of the shape, known a priori at learning time, and unadapted to the physical nature of the problem.
	
	In this work, we propose a new method, that alleviates the problems cited above. We introduce a new type of deep convolutional neural networks able to perform convolutions over a geodesic surface described by a mesh, in our case the very shape we are aiming to optimize, and train such a Geodesic Convolutional Neural Network (GCNN) to predict the performance measure of interest (e.g. CX, drag, lift). We then use the GCNN predictor as a new differentiable black-box surrogate model that maps a shape design to the objective. We optimize this approximation using gradient descent on the shape design itself. We hence obtain a new optimisation method, which is several orders of magnitude faster than previous approaches, non parametric, and can take into account additional shape constraints in a very flexible way.
}


\section{Introduction}

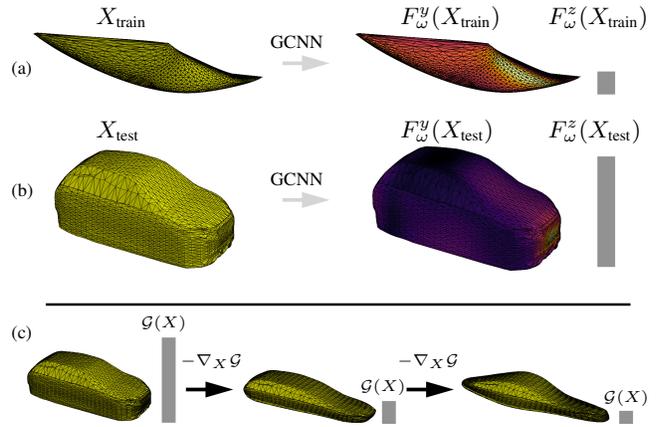
\begin{figure}[ht!]
	\centering
	\begin{overpic}[ width=85mm, trim={-3 70 3 0},clip]{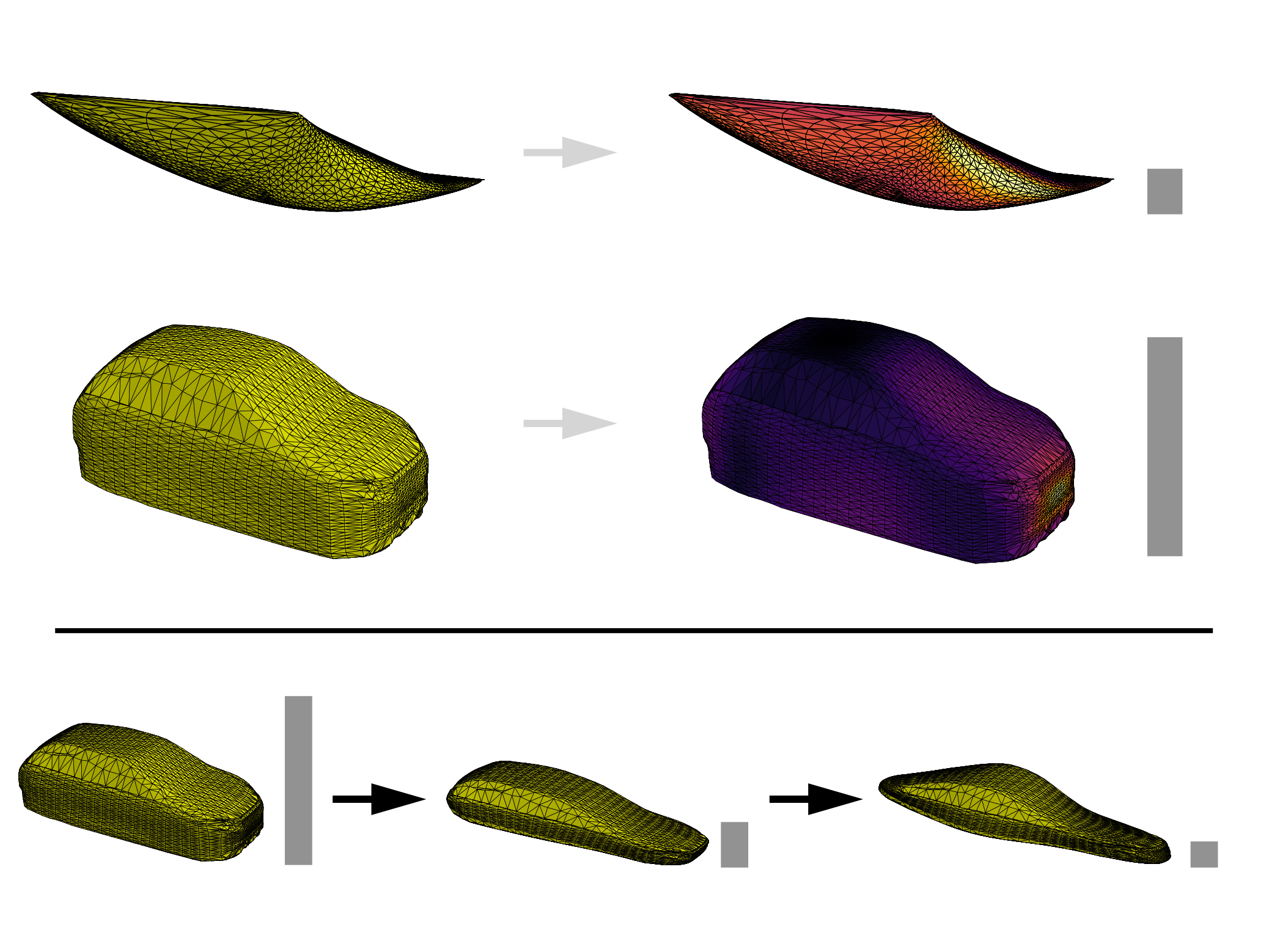}
		\put(62,22){\tiny {$-\nabla_X  \mathcal G $}}
		\put(144,22){\tiny {$-\nabla_X  \mathcal G $}}
		\put(96,90){\scriptsize {GCNN}}
		\put(96,142){\scriptsize {GCNN}}
		\put(30,151){\footnotesize {$X_{\text{train} }$}}
		\put(30,108){\footnotesize {$X_{\text{test} }$}}
		\put(145,151){\footnotesize {$F^y_{\omega} (X_{\text{train} })$}}
		\put(145,108){\footnotesize {$F^y_{\omega} (X_{\text{test} })$}}
		\put(201,151){\footnotesize {$F^z_{\omega} (X_{\text{train} })$}}
		\put(201,108){\footnotesize {$F^z_{\omega} (X_{\text{test} })$}}
		\put(47,36){\tiny {$\mathcal G (X)$}}
		\put(130,12){\tiny {$\mathcal G (X)$}}
		\put(222,8){\tiny {$\mathcal G (X)$}}
		\put(-2,32){\scriptsize {(c)}}
		\put(-2,86){\scriptsize {(b)}}
		\put(-2,132){\scriptsize {(a)}}
	\end{overpic}
	\vspace{-5mm}
	\caption{Car shape optimization. (a) We train a GCNN to emulate a CFD simulator and predict from the shape on the left pressures shown as colors in the middle and drag shown as a gray vertical bar on  the right.  (b) Convolutional layers enable the GCNN to make accurate predictions on previously unseen shapes, such as those of cars. (c) We use the GCNN to express drag as a differentiable function of the mesh vertex positions. Finally, we minimize it under constraints to produce the rightmost shape.}
	\vspace{-4mm}
	\label{fig:teaser}
\end{figure}

\comment{
\begin{figure}[ht!]
	\centering
   \begin{overpic}[ width=85mm, trim={-3 120 3 0},clip]{figures/fig_teaser/teaser_lowres.png}
   	\put(68,32){\tiny {$-\nabla_X  \mathcal G $}}
   	\put(150,32){\tiny {$-\nabla_X  \mathcal G $}}
   	\put(73,89){\scriptsize {GCNN}}
   	\put(73,134){\scriptsize {GCNN}}
   	\put(30,142){\footnotesize {$X_{\text{train} }$}}
   	\put(30,108){\footnotesize {$X_{\text{test} }$}}
   	\put(110,142){\footnotesize {$F^y_{\omega} (X_{\text{train} })$}}
   	\put(110,108){\footnotesize {$F^y_{\omega} (X_{\text{test} })$}}
   	\put(195,142){\footnotesize {$Y_{\text{train} }$}}
   	\put(195,108){\footnotesize {$Y_{\text{test} }$}}
   	\put(10,46){\scriptsize {$\mathcal G (X) = 80.21$}}
   	\put(102,46){\scriptsize {$\mathcal G (X) = 4.91$}}
   	\put(185,46){\scriptsize {$\mathcal G (X) = 4.71$}}
   	\put(-2,41){\scriptsize {(c)}}
   	\put(-2,86){\scriptsize {(b)}}
   	\put(-2,132){\scriptsize {(a)}}
   	\end{overpic}
	\caption{ \pf{Car shape optimization. (a) We train a GCNN to emulate a CFD simulator and predict from the shape on the left pressures shown as colors in the middle and drag shown as a vertical bar on  the right.  (b) Convolutional layers enable the GCNN to makers accurate predictions on previously unseen shapes, such as those of cars. (c) We use the GCNN to express drag as a differentiable function of the mesh vertex positions. Finally, we minimize it under constraints to produce the rightmost shape.} }
	\label{fig:teaser}
\end{figure}
}


Optimizing the aerodynamic or hydrodynamic properties is key to designing shapes, such as those of aircraft wings, windmill blades, hydrofoils, car bodies, bicycle shells, or submarine hulls. However, it remains challenging and computationally demanding because existing Computational Fluid Dynamics (CFD) techniques rely either on solving the Navier-Stokes equations or on Lattice Boltzmann methods. Since the simulation must be re-run each time an engineer wishes to change the shape, this makes the design process slow and costly. A typical engineering approach is therefore to test only a few designs without a fine-grained search in the space of potential variations. 


This is a severe limitation and there have been many attempts at overcoming it, but none has been entirely successful yet. Most current algorithms rely on combining evolutionary algorithms with heuristic local search~\cite{Orman16} and complex adjoint methods~\cite{Allaire15, Gao17}, which requires rerunning a simulation at each iteration step and therefore remains costly. A classical approach to reduce the computational complexity is to use Gaussian Process (GP) regressors trained to interpolate the performance landscape  given a low dimensional parametrization of the shape space.  This interpolator is then used as a proxy for the true objective to speed-up the computation, which is referred to as Kriging in the CFD literature~\cite{Toal11,Xu17}.  However, those regressors are only effective for shape deformations that can be parameterized using relatively few parameters~\cite{Bengio06} and  their performance therefore hinges on a well-designed parameterization. Furthermore, the regressors are specific to a particular parameterization and pre-existing computed simulation data using different ones cannot be easily leveraged. 

By contrast, we propose an approach to optimizing the aerodynamic or hydrodynamic properties of arbitrarily complex shapes by training Geodesic Convolutional Neural Networks (GCNNs)~\cite{Monti16} to emulate a fluidynamics simulator.  More specifically, given  a set of generic surfaces parametrized as meshes,  we train a GCCN to predict their aerodynamic characteristics, as computed by standard CFD packages, such as XFoil~\cite{Drela89} or Ansys Fluent~\cite{Ansys11}, which are then used to write an objective function.  Since this function is differentiable with respect to the vertex coordinates, we can use it  in conjunction with a gradient-based optimization to explore the shape space much faster and thoroughly than was possible before, with a much better accuracy and without putting undue restrictions on the range of potential shapes. Since performing convolutions on an arbitrary mesh is much slower than on a regular grid, the key to making this approach practical is remeshing the original shape using a cube or poly-cube map~\cite{Tarini04}. It makes it possible to perform the GCNN computations on GPUs instead of CPUs, as for normal CNNs.

In short, our contribution is a new surrogate model method for shape optimization. It does not rely on handcrafted features and can handle shapes that must be parameterized using many parameters. Since it operates directly on the 3D surfaces, an added benefit is that it can leverage training data produced using any kind of parameterization and not only the specific one used to perform the shape optimization. Fig~\ref{fig:teaser} illustrates the process. The training shapes can be very different from the target one, which gives our system flexibility.

We demonstrate that our method outperforms GP approaches, widely used in the industry, both in terms of regression accuracy and optimization capacity. Not only do we improve upon GP optimization by 5\% to 20\% for a lift maximization task on 2D NACA airfoils involving few parameters, but we can also deliver results on fully 3D shapes for which our baselines perform poorly. 


\comment{

One way to reduce the computational complexity is to rely on low-dimensional shape models, which constrains the shape to remain in a low-dimensional manifold and therefore ignores potential solutions away from it. One popular class of methods relies on aThese so-called {\it Meta-Model} or {\it Surrogate} techniques are effective but are also limited to shapes that can be parameterized in terms of a limited number of parameters. Also the parametrization has to be know a-priori to build a training database inside these bounds. Finally, this approach focuses on interpolating the objective function without regards to the complex physical phenomena that make take place as the shape simulated shape changes, which limits the method's generalization abilities.

By keeping the CNN parameters fixed, we can use a standard projected gradient descent optimization scheme to optimize the predicted aerodynamic performance with respect to the variable input shape and taking into account additional shape constraints. Note that during the optimization process , the CNN can be trained further, leveraging our \textit{black-box} simulator and generating new samples around the currently explored shapes in order to increase the accuracy of the predicted aerodynamic performance.

For every time step, each proposed airfoil shape has to be evaluated using Computational Fluid Dynamic simulators in a \textit{black-box} fashion. 
Such computations are very expensive and hence limit the overall efficiency of the optimization algorithms.
Furthermore, since we don't have access to the gradients of the objective with respect to the shape, many of its local deformations may have to be evaluated to find a good direction of descent. 
Note that, in order to reduce the complexity of the optimization space, such methods need parametric models for airfoil shapes, which are inherently restricted to stay in a low dimensional manifold inside the high dimensional space of all possible shapes.

Another class of methods often used to in practice, called Meta-Models or Surrogate ones, Such methods require to train a parameterization specific aerodynamic regressor as they don't perform well on relatively high-dimensional shape spaces, making de-facto impossible to leverage on a database of classified shapes and requiring careful choice of the parameters before the optimization starts. Furthermore, this regression method does not leverage on the physical, locally invariant nature of the phenomenons that we it is trying to predict.

In this work, we propose a new approach to aerodynamic shape optimization based on Geodesic Convolutional Neural Network (CNN) proxies, which alleviates all of the pre-cited problems. Geodesic Convolutions belong to the class of recently developed Geometric Deep-Learning methods, which extend the standard Convolutional Neural Network paradigm to handle non-Euclidean structures such as graphs or meshes. The main motivation behind this method is to interpolate locally the features from neighbouring nodes in before making the updating the value at a node.

We first train a Geodesic CNN which, from a non-parametric discretized mesh describing an aerodynamic shape, predicts its polars: the dimensionless pressure coefficient $C_p$ and drag coefficient $C_d$. In particicular our CNN takes as input a generic surface mesh $\mathcal{M} = \{(V,E) \}$, defined by the coordinate of its vertices and the associated connectivity graph, and uses geodesic convolutions to regress the pressure coefficients at each vertex of the mesh. It finally uses a Fully-Connected layer to predict the drag coefficient. Note that even if, motivated by practical applications, we considered the regression of polars $C_p \in \mathbb{R} ^ {| V |} $ and $C_D \in \mathbb{R} $ our framework can support the regression of any other polar of interest.

Compared to previous works on surrogate methods for shape optimization, our main innovation is the introduction of the Geometric Convolutional Neural Network. Hence, our method doesn't require any arbitrary parametrization of the input, making it more versatile and general. Furthermore, we gain the ability to optimize the surrogate with gradient based methods and keeping the computational cost low, even in 3 dimensions.

}


\section{Related Work}
\label{sec:related}

There is a massive body of literature about  fluid simulation techniques. Traditional ones  rely on numerical discretization of the Navier-Stokes Equations (NSE) using finite-difference, finite-element, or finite volume methods~\cite{Quarteroni09,Skinner18}. Since the NSE are highly non-linear, the space has to be very finely  discretized for good accuracy, which tends to make such methods computationally expensive.  In some cases, approaches such as the \textit{Lattice-Boltzmann Method} (LBM)~\cite{McNamara88}, which simulates streaming and collision processes across individual fluid particles to recover the global behavior as an emergent property, can be more accurate, at the cost of being even more computationally expensive~\cite{Xan11}. 

All the above-mentioned techniques approximate the  fluodynamics for a fixed physical shape and without changing it. A common engineering practice is therefore to first use them to compute the characteristics of a few hand-designed shapes and then to pick the one giving the best results. In this paper, our focus is on using a Deep Learning approach to automate this process by casting it as a gradient descent minimization. In the remainder of this section, we therefore review existing approaches to shape optimization in the CFD context and then discuss current uses of Deep Learning in that field. 

\subsection{Shape Optimization}
\label{sec:shapeOpt}

A popular and relatively easy to implement approach to shape optimization relies on genetic algorithms~\cite{Gosselin09} to explore the space of possible shapes. However, since genetic algorithms require many evaluation of the fitness function, a naive implementation would be inefficient  because each one requires an expensive CFD simulation.

This can be avoided using Adjoint Differentiation~\cite{Allaire15, Gao17} instead. It involves approximating the solution of a so-called {\it adjoint} form of the NSE to compute the gradient of the fitness function with respect to the 3D simulation mesh parameters. This allows the use of gradient-based optimization techniques but is still very expensive because it requires a new simulation to be run at each iteration~\cite{Alexandersen16}. 

As a result, there has been extensive research in ways to reduce the required number of evaluations of the fitness function. One of the most widely used one relies either on Gaussian Processes (GPs, \citealp{Rasmussen06}), which is known as {\it kriging} in the CFD community~\cite{Jeong05,Toal11,Xu17}, or on Artificial Neural Networks (ANN, \citealp{Gholizadeh11,Lundberg15,Preen15}) to compute {\it surrogates} or {\it meta-models} that approximate the fitness function while being much easier to compute. Both approaches can be used either to simply speed-up genetic algorithms~\cite{Ulaganathan13} or to find an optimal trade-off between exploration and exploitation, using the confidence bounds provided by the GPs to optimize a multi-objective Pareto front. However, these methods all depend on handcrafted shape parameterizations that can be difficult to design.  Furthermore this makes it necessary to retrain the regressor for each new scenario in which the parameterization changes. 


\subsection{Deep Learning}
\label{sec:deepLearning}

As in many other engineering fields, Deep Learning was recently introduced in CFD. For example, Neural Networks are used by~\citet{Tompson16} to accelerate Eulerian fluid simulations. This is done by training  CNNs to approximate the solution of the the discrete Poisson equation, which is usually the most time demanding step in an Eulerian fluid simulation pipeline.  Along similar lines, 3D CNNs are used by~\citet{Guo16}  to  directly  regress the fluid velocity field at every point of the space given an implicit surface description of the target object. 

These two methods demonstrate the potential of Deep Learning to speed-up and reproduce fluid simulations. However, they rely on 3D CNNs which have a large memory footprint and are extremely computationally demanding whereas the object of interest is intrinsically represented as  2D manifold. This can be mitigated by running the simulations over a coarse space discretization, which degrades the accuracy. By contrast, our proposed approach directly runs on a surface mesh representation of the object, which enables us to use one less dimension and thus considerably reduce the computing requirements.



\section{Regression of physical quantities}
\label{sec:formulation}

We define the set of meshes  $\mM  \subset \mathbb{R}^{3N} \times \{0, 1\}^{N \times N} $
and 
$
\bM = \left(\bX, \mE\right) \in \mM\;,
$
a mesh as being a pair composed of locations of vertices, and their connectivity.

Let us assume we are given a set of meshes
$
\bM_m = (\bX_m, \mE_m), \ m = 1, \dots, M,
$
and let us further assume that for each one, we ran a CFD solver to compute both a corresponding vector of local physical quantities $\bY_m \in \bR^{N}$, one for each vertex,  along with a global scalar value $Z_m \in \bR$. Concretely,  $\bY_m$ could be the air-pressure field along a plane's wing and $Z_m$ the total drag force it generates. From these values we can infer a performance score, such as Lift-to-Drag ratio for a wing, using a differentiable mapping.

Given $M$ such triplets $(\bM_m,\bY_m, Z_m)$, we want to train a regressor $F_{\omega} : \mM \rightarrow \bR^N \times \bR$ such that
\begin{equation}
 F_{\omega}(\bM_m)  = \left(F^y_{\omega}(\bM_m),F^z_{\omega}(\bM_m) \right) \approx \left(\bY_m, Z_m\right), \label{eq:regressor}
 \end{equation}
where  $\omega$ comprises the trainable parameters, which will be optimized to minimize our training loss
\begin{equation}
\hspace*{-0.5em}{\cal L}(\omega) \!=\! \sum_m  \| F^y_{\omega}(\bM_m) \!-\! \bY_m \|^2 + \lambda \left(F^z_{\omega}(\bM_m) \!-\! Z_m\right)^2,\label{eq:fitting}
\end{equation}
%
where $\lambda$ is a scaling parameter that ensures that both terms have roughly the same magnitude.

\subsection{From Geodesic to Cube-Mesh CNNs}
\label{sec:geodesicCNN}

Standard CNNs implicitly rely on their input, images usually, having a regular Euclidean geometry. The neighborhood relationship between pixels encodes their distance. 
While such a regularity is true for images, it is not for surface meshes. To operate on such an input, one therefore should use Geodesic CNNs (GCCNs) such as those described by~\citet{Monti16} instead. These explicitly account for the varying geodesic distances between vertices when performing convolutions. 

However, the structure of the input -- the fact that it is organised as a tensor where adjacent elements are neighbours in the physical world--, is  also central to the efficient use of modern computational hardware and results in speed-up of several orders of magnitude. 
For our application, the lack of structure of arbitrary surface meshes -- often composed of triangles -- would slow down the computation and prevent effective use of the GPUs for a naive implementation of GCNNs.
In this section, we first introduce GCNNs which help overcome the first difficulty and then show how we can remesh the surface into a quad-mesh to tackle the second one.


\paragraph{Geodesic CNNs.}
We describe the geometric convolution operation that is used at each vertex, where a mixture of gaussians is used to interpolate the features computed at neighbouring vertices into a common predefined basis.

Let us consider a signal $f = (f^1, \ldots, f^N)$ defined at each one of the $N$ vertices ${\bX^i}_{1 \leq i \leq N}$ of mesh  $\bM$.
For each $i$, let $\mN^i = \{ j: \mE(i, j) = 1 \}$, that is the set of indices $j$ such that ${\bX^i}$ and ${\bX^j}$ are neighbors.

\comment{Note that  $\mN^i$ can either be limited to immediate neighbors or include more distant ones, depending on the application.}  Let $K$, where $K=32$ in all our experiments, be a predefined number of gaussian parameters $\alpha_k \in \bR^2$, $\Sigma_k \in \bR^2$, which are vertex independent. We can now define an interpolation operator $D_k$  over the mesh vertices by writing 
\begin{eqnarray}
\hspace*{-1em}(D_kf)^i = \sum \limits_{j \in \mN^i} & \!\!\!f^j & \hspace{-2mm} \exp\left( \frac{-(\rho(\bX^i,\bX^j) - {\alpha_k}_\rho)^2}{{\Sigma_k}_\rho}\right)  \nonumber \\
                                                    &           & \exp\left( \frac{-(\theta(\bX^i,\bX^j) - {\alpha_k}_\theta)^2}{{\Sigma_k}_\theta}\right) \label{eq:interpolate}, \ 
\end{eqnarray}
where $\rho(\cdot)$ and $\theta(\cdot)$ are relative geodesic coordinates. This makes it possible to define the convolution of $f$ by a filter $g$ over the mesh as 
\begin{equation}
f \star g = \sum \limits_{k \in 1, \dots, K} g_kD_k f.
\label{eq:conv}
\end{equation}
This operator can then be used as a building block for a convolutional architecture. The learning phase involves a gradient-descent based optimization of the convolutional function parameters $g_k$, as in standard CNNs. The values of $\alpha_k$ and $\Sigma_k$ can be set manually and kept fixed during training~\cite{Kipf16}. However, it is more effective to learn them~\cite{Monti16}. 

Unfortunately, the meshes  we must deal with are typically large and a naive implementation of the GCNNs would be prohibitively expensive because the convolution of Eq.~\ref{eq:interpolate} lacks the structure that would make it easy to implement on a GPU, forcing the use of the CPU, which is much slower. In theory, this could be remedied by storing the exponential terms of Eq.~\ref{eq:interpolate} in {\it adjacency matrices}, as by~\citet{Monti16}. However, densely representing those matrices, would not be practical because it would be too large to fit on the GPU for real-world problems. A sparse representation would solve the memory problem but would be equally impractical because, unlike by~\citet{Monti16}, the geodesic distances change at every iteration and have to be recomputed. Filling these new values into a sparse representation that fits on the GPU  would also be very slow. 

\paragraph{Cube-Mesh mapping.} 


\vspace{-0.3cm}
\begin{figure}[t!]
	\captionsetup[subfigure]{labelformat=empty}
	\centering
		\begin{overpic}[width=80mm, trim={0 0 0 50},clip]{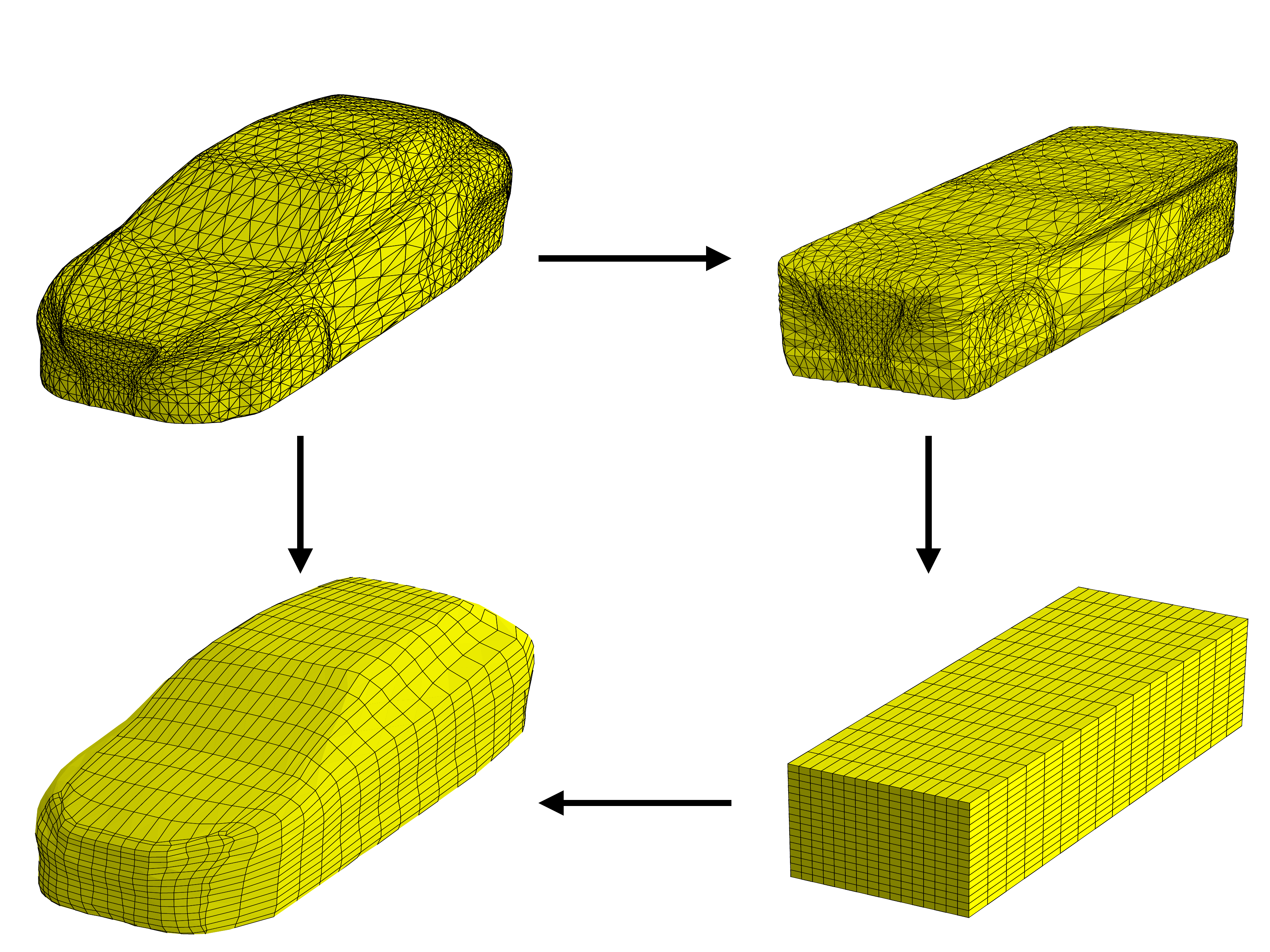}
		\put(12,145){\footnotesize{(a)}}
		\put(152,145){\footnotesize{(b)}}
		\put(152,57){\footnotesize{(c)}}
		\put(12,57){\footnotesize{(d)}}
		\put(15,84){\footnotesize{quad}}
		\put(5,74){\footnotesize{re-meshing}}
		\put(187,84){\footnotesize{grid}}
		\put(176,74){\footnotesize{extraction}}
		\put(97,140){\footnotesize{volumetric}}
		\put(95,130){\footnotesize{deformation}}
		\put(97,33){\footnotesize{projection}}
		\end{overpic}
	\vspace{-3mm}
	\caption{Remeshing. (a) A triangular mesh is morphed into a \textit{pseudo} cube (b) aligned with its principal axes by iteratively applying volumetric deformations~\cite{Gregson2011}. A semi-regular quad-mesh (c) is then defined on the pseudo-cube  and we project its vertices onto the original mesh to create the regular tesselation of the original shape shown in (d). }
	\vspace{-2mm}
	\label{fig:cubemapping}
\end{figure}

To overcome this difficulty, we propose to exploit the properties of cube and poly-cube maps~\cite{Tarini04}, which allow the remeshing of 3D shapes homotopic to spheres such as those of cars or bicycles shells as semi-regular quad-meshes with few irregular vertices, a uniform tessellation density and well-shaped quads~\cite{Bommes13}. Fig.~\ref{fig:cubemapping} illustrates this process. After remeshing, we can perform the convolutions of Eqs.~\ref{eq:interpolate} and~\ref{eq:conv} for all regular vertices using GPU acceleration. However, a few irregular vertices are unavoidable for most shapes, according to index theory~\cite{Bommes12}.  We handle them as special cases, at the cost of a small approximation described in the supplementary material. 

In practice, we have in memory one ``image'' per face of the bottom-right box of Fig.~\ref{fig:cubemapping}, and  each pixel of the said images has three channels corresponding to the 3d coordinates of that point in the remeshed shape, which is used to compute the values of $\rho()$ and $\theta()$, which can now be done in a GPU-friendly manner.

\subsection{Regressor Architecture}
\label{sec:archi}


\begin{figure*}[ht!]
	\centering
	\begin{overpic}[width=160mm, trim={-10 210 0 210},clip]{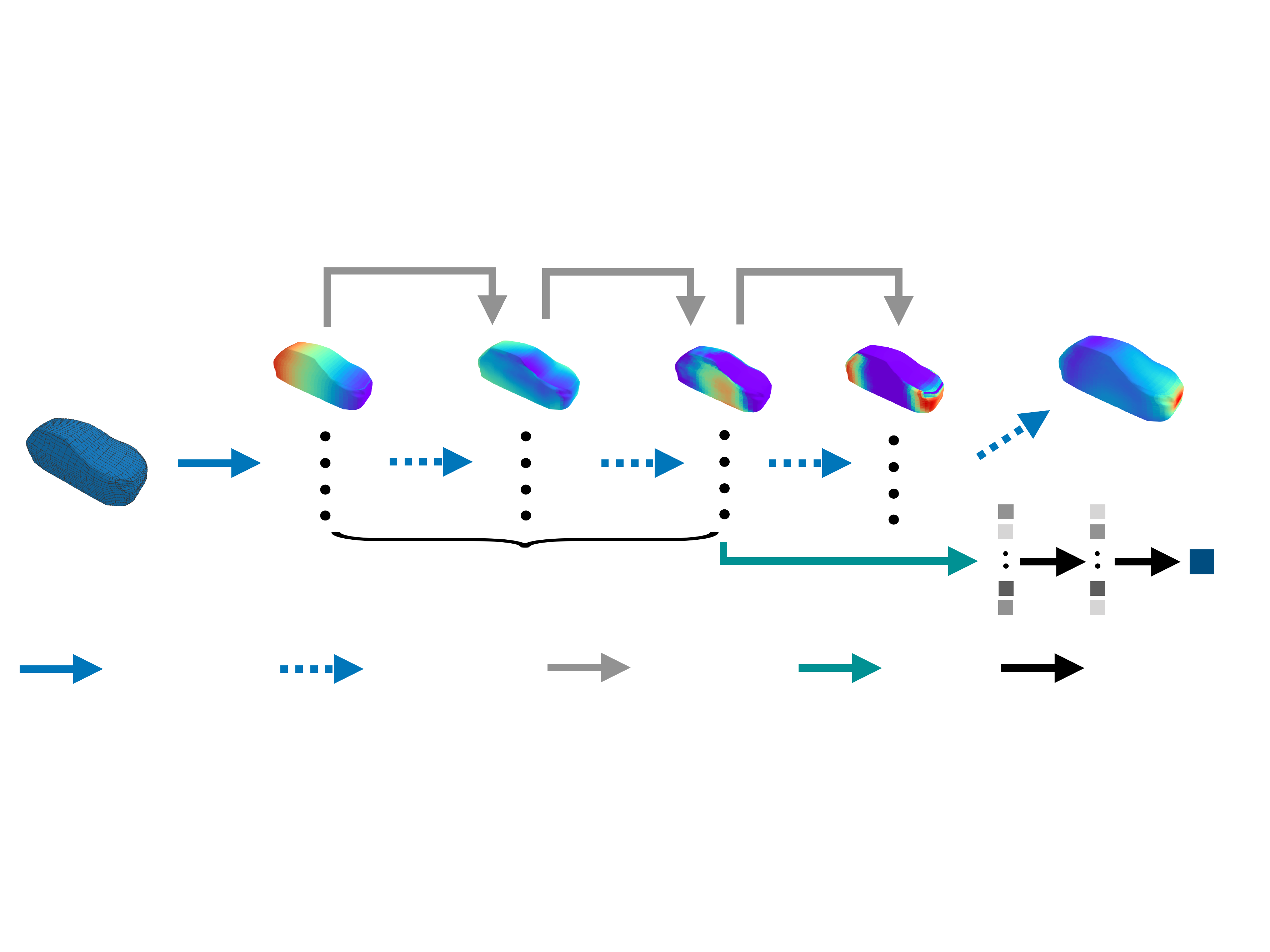}
		\put(32,100){\normalsize{$X$} }
		\put(410,130){\normalsize{$F^y_{\omega}$} }
		\put(426,57){\normalsize{$F^z_{\omega}$} }
		\put(185,38){\normalsize{$F^0_{\omega}$} }
		\put(42,11){\footnotesize{Atrous geo } }
		\put(54,4){\footnotesize{conv} }
		\put(135,11){\footnotesize{Atrous geo } }
		\put(136,4){\footnotesize{conv block } }
		\put(232,11){\footnotesize{Residual } }
		\put(230,4){\footnotesize{connection } }
		\put(321,11){\footnotesize{Mean } }
		\put(319,4){\footnotesize{pooling } }
		\put(390,11){\footnotesize{Fully connected } }
		\put(409,4){\footnotesize{layer } }
		
		\end{overpic}
        \vspace{-4mm}
	\caption{The architecture of our geometric CNN.  The vertex coordinates of the quad-mesh of Fig.~\ref{fig:cubemapping} are fed to a convnet that produces a feature map, which is itself fed to both another convnet and a fully connected one. The first outputs a vector of pressure values and the second a scalar drag value. Using the same features for both prevents overfitting.}
	\vspace{-2mm}
	\label{fig:architecture}
\end{figure*}

Recall from Eq.~\ref{eq:regressor} that our regressor must predict a vector of local values $Y=F^y_{\omega}(\bM)$ and a global scalar value $Z=F^z_{\omega}(\bM)$ given a mesh $\bM$, whose shape and topology are defined by a vector of 3D vertex coordinates $\bX$ and a set of edges $\mE$. Our Network architecture  $F_{\omega}$, depicted in Fig.~\ref{fig:architecture} includes a common part $F^0_{\omega}$  that processes the input and feeds two separate branches $F^y_{\omega}$ and $F^z_{\omega}$, which respectively predict $\bY$ and $Z$. $F^0_{\omega}(\bM)$ is a feature map $\in \bR^{N\times k}$, where $k$ is the number of features for each one of the $N$ vertices. $F^y_{\omega}$  is another Geodesic-Convolutional branch that returns $\bY  \in \bR^N$ while $F^z_{\omega}$ regresses the scalar value $Z \in \bR$ by average pooling followed by two dense layers. Effectively, our  shared convnet  $F_{\omega}$ therefore takes as input this vector $\bX$ to predict the desired vector $Y$ and scalar value $Z$.

Interestingly, as shown in the experiments, we noticed that by learning to predict more physical quantities than actually needed, through additional branches, as by~\citet{Caruana97,Ramsundar15} 
we favour the emergence intermediate-level features that are more robust and less overfitting prone. This observation hints that our architecture is actually learning physical phenomenons and not only interpolating the output.

As discussed above, all these operations could be implemented using geodesic nets that operate directly on $\bM$ and perform convolutions such as those of Eq.~\ref{eq:interpolate}, but this would be slow. In practice, we first map a reference shape on a cube such as the one depicted at the bottom right of Fig.~\ref{fig:cubemapping} to obtain a semi-regular quad-remeshing. Then, $\bX$ becomes the set of 3D coordinates assigned to each vertex of the resulting regular vertex grid, which we then use to reconstruct either identical or modified versions of the 3D shape, such as the one shown at the bottom left of Fig.~\ref{fig:cubemapping}. 
To increase the receptive field of our convolutions without needlessly increasing the number of parameters or reducing the resolution of our input, we use dilated convolutions~\cite{Yu16b}, along with several convolutional blocks with pass-through {\it residual} connections~\cite{He16}.






\section{Shape Optimization}
\label{sec:optim}

Once trained, the regressor $F_{\omega}$ can be used as a surrogate model or proxy to evaluate the effectiveness of a particular design. Not only is it fast, but it is also {\it differentiable} with respect to the $\bX$ coordinates that control the shape. We can therefore use a gradient-based technique to maximize the desirable property while enforcing design constraints, such as the fact that a bicycle shell must be wide enough to accommodate the rider, a car  must contain a pre-defined volume for the engine and passengers, or a plane wing should be thick enough to have the required structural rigidity. 

Formally, this can be expressed by treating $F_{\omega}$  as a function of $\bX$ and looking for 
%
%
\begin{equation}
\bX^\ast = \argmin_\bX \mG \left( F_{\omega}(\bX) \right)  \text{\ \ \  s.t\ \ \ } \mC(\bX) \leq 0, \label{eq:optim} 
\end{equation}
where $\mG$ is a {\it fitness function}, such as the negative Lift-to-Drag ratio in the case of a wing or simply the drag in the case of the car, and $\mC$ represents the set of constraints that a shape must satisfy to be feasible. 

\subsection{Projected Gradient Descent}
\label{sec:projGrad}

\begin{figure}[ht!]
	\centering
	\vspace{-2mm}
	\begin{overpic}[ width=85mm, trim={0 610 240 0},clip]{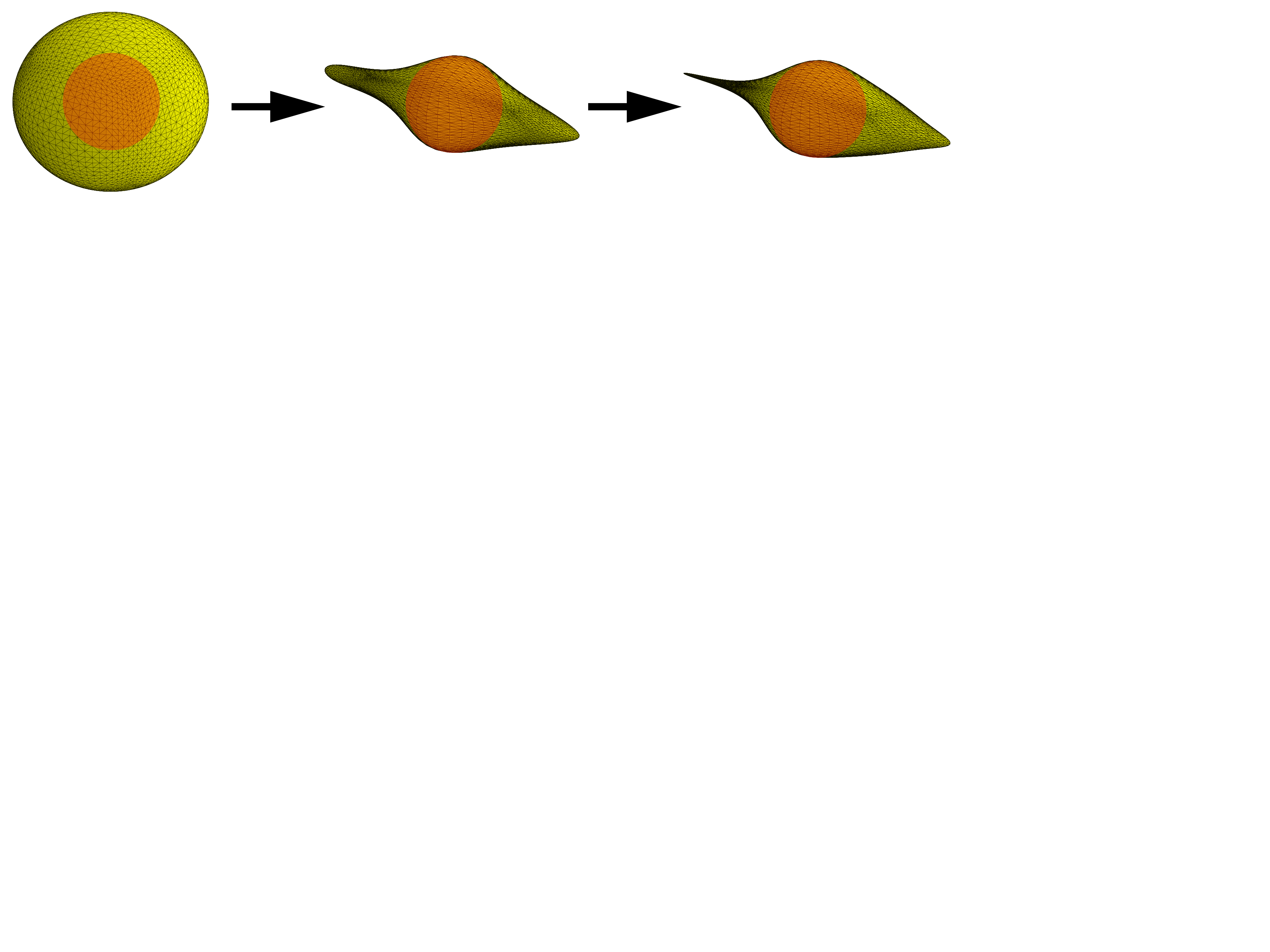}
		\put(55,29){\tiny {$-\nabla_X  \mathcal G $}}
		\put(142,29){\tiny {$-\nabla_X  \mathcal G $}}
	\end{overpic}
	\vspace{-5mm}
	\caption{Minimizing the drag of the initially spherical surface under the constraint it must contain the smaller red sphere.}
	\vspace{-2mm}
	\label{fig:sphere}
\end{figure}

To solve the minization problem of Eq.~\ref{eq:optim}, we use a projected version of the popular ADAM algorithm~\cite{Kingma15}: at the end of each iteration, we check if the $\bX$ still is a feasible shape. If not, we project it to closest feasible point. To do this effectively, we need to compute the Jacobians 
\begin{equation}
\nabla_{\bX} \mG \left(F_{\omega}(\bX) \right)\; \mbox{and } \nabla_{\bX}  \mC(\bX), 
\end{equation}
which we do using the standard chain rule through automatic differentiation. Fig.~\ref{fig:sphere} depicts such a minimization. 


\subsection{Parametrization}
\label{sec:parametrization}

Given the size of the meshes we deal with, the optimization problem of Eq.~\ref{eq:optim} is a very large one, which traditional methods such as Kriging~\cite{Toal11,Xu17} cannot handle. To reduce the problem's dimensionality and make comparisons possible,  we introduce parametric models. More specifically, when the shape and its pose are controlled by a small number of parameters $\bC$, we can write the vertex coordinates as a differentiable function of these parameters $\bX(\bC)$ and reformulate the minimization problem of Eq.~\ref{eq:optim} as a minimization with respect to $\bC$. For example, wings have long been described in terms of their NACA-4digits parameters~\cite{Jacobs48}, which are three real numbers representing a family of shapes known to have good aerodynamic properties. These numbers correspond to the maximum camber, its location, and the wing thickness. 

By contrast to Kriging, the performance of our approach increases with the number of parameters, which we will demonstrate in the result section by using 18 parameters instead of only the 3 NACA ones for airfoil profiles. To further increase flexibility, we could replace the parametric models by the Laplacian parameterization introduced by~\citet{Ngo16}, as demonstrated in the 2D case in the supplementary material. It expresses all vertex coordinates as linear combinations of those of a small number of control vertices. Thus, the objective function will become a differentiable function of the positions of a subset of mesh vertices. Our approach will therefore directly apply, which would let us adjust the model's complexity as needed by adding or removing control points. 

\comment{

\paragraph{Laplacian Parameterization}

When more flexibility is required, we rely on the Laplacian parameterization introduced by~\citet{Ngo16}, which expresses all vertex coordinates as linear combinations of those of a small number of control vertices. More specifically, we write 
\begin{equation}
\bX = \bP \bC, 
\end{equation}
where $\bP$ is a matrix that is computed once for a reference shape and $\bC$ are the 3D coordinates of a subset of the mesh vertices that act as control points. $F_{\omega}$ then becomes a function of $\bC$. The regularity of the shape can then be controlled by increasing or decreasing the number of control points. In the limit, all mesh vertices can become control vertices and the matrix $\bP$ reduces to the identity. 

\PF{You don't want to mention the regularization term? We might have to once Edoardo implements the control points.} \PB{It was just to shorten a bit the text but, yes, it should be added if necessary and if we use it.}
}

\comment{
\subsection{Laplacian Parameterization}
\label{sec:laplacian}

Given that the meshes we deal with, the optimization problem of Eq.~\ref{eq:optim} is a very large one. To reduce its dimensionality,  we use the Laplacian parameterization introduced in~\ref{Ngo16} that expresses all vertex coordinates as linear combinations of those of a small number of control vertices. More specifically, we write 
\begin{equation}
\bX = \bP \bC, 
\end{equation}
where $\bP$ is a matrix that is computed once for a reference shape and $\bC$ are the 3D coordinates of a subset of the mesh vertices that act as control points. This representation yields a comparatively-low dimensional and linear representation of complex deformations with a minimal loss of accuracy while being completely rotationally invariant, unlike PCA style representations. 

$F_{\omega}$ then becomes a function of $\bC$. The regularity of the shape can then be controlled by increasing or decreasing the number of control points. In the limit, all mesh vertices can become control vertices and the matrix $\bP$ reduces to the identity. Additional regularity can be imposed by writing the regularization term of Eq.~\ref{eq:optim} as
\begin{equation}
\mR (\bC ) \propto \| \bA \bP \bC  \|^2 =  \| \bA \bX  \|^2\; , 
\end{equation}
where $\bA$ also is a precomputed matrix. This term tends to favor shapes in which the curvature remains relatively similar to that of the reference shape used to compute the matrices $\bP$ and $\bA$. 
}

\subsection{Online Learning}
\label{sec:online}

We start from of an initial set of  random shapes on which we run a full simulation to generate the triplets $(\bM_i,\bY_i, Z_i)$. We then use it to train $F_{\omega}$  by minimizing the loss of Eq.~\ref{eq:regressor}.

If the database used to train the network is not representative enough, $\bX$ can drift away from regions of the shape space where  our proxy provides a good approximation. Since performing even a single simulation is much slower than running many ADAM optimization steps, we alternate between the following two-steps.
\vspace{-2mm}
\begin{enumerate}

 \item We run project gradient steps as discussed above using the current $F_{\omega}$ GCNN regressor until convergence. 
 
 \item We run a new simulation for the obtained shape, add this new sample to the training set and fine tune the $F_{\omega}$ GCNN regressor with this new training sample. 
\end{enumerate}
\vspace{-2mm}
Note that in an industrial setting, the randomly chosen set of initial samples could be replaced by all the shapes that have been simulated in the past. Over time, this would result in an increasingly competent proxy that would require less and less re-training.


%


\section{Experimental Results}
\label{sec:results}

In this section, we evaluate our proposed shape optimization pipeline. It is designed to handle 3D shapes but can also handle 2D ones by simply considering the 2D equivalent of a surface mesh, which is a discretized 2D contour. We therefore first present results on 2D airfoil profiles, which have become a {\it de facto} standard in the CFD community for benchmarking shape optimization algorithms~\cite{Toal11,Orman16}. We then use the example of car shapes to evaluate our algorithm's behavior in the more challenging 3D case. We implemented our deep-learning algorithms in TensorFlow~\cite{Tensorflow} and ran them on a single \texttt{Titan X Pascal} GPU. 

We will quantify the accuracy of various regressors in terms of the standard $L^2$ mean percentage error over a test set $\mathcal S _v $, that is, 
\begin{equation}
\mathcal A _{y} = 1.0 - \bE_{n \in S_v} [ \tfrac{\| y_n - \hat y_n\|_2 }{\|  y_n \| _2}] \; ,
\label{eq:accuracy}
\end{equation}
where $y$ denotes either a ground truth local quantity $\bY_i$ or the global one $Z_i$. In turn, $\hat y$ denotes the corresponding predictions $\mF^y(\bX_i)$ or $\mF^z(\bX_i)$.

\subsection{2D Shapes - Airfoils and Hydrofoils}
\label{sec:2Dflow}


\begin{figure}[thbp]
\begin{center}
\begin{tabular}{cc}
	\begin{overpic} [scale=0.30,unit=1mm,trim={60 -15 60 -15},clip]{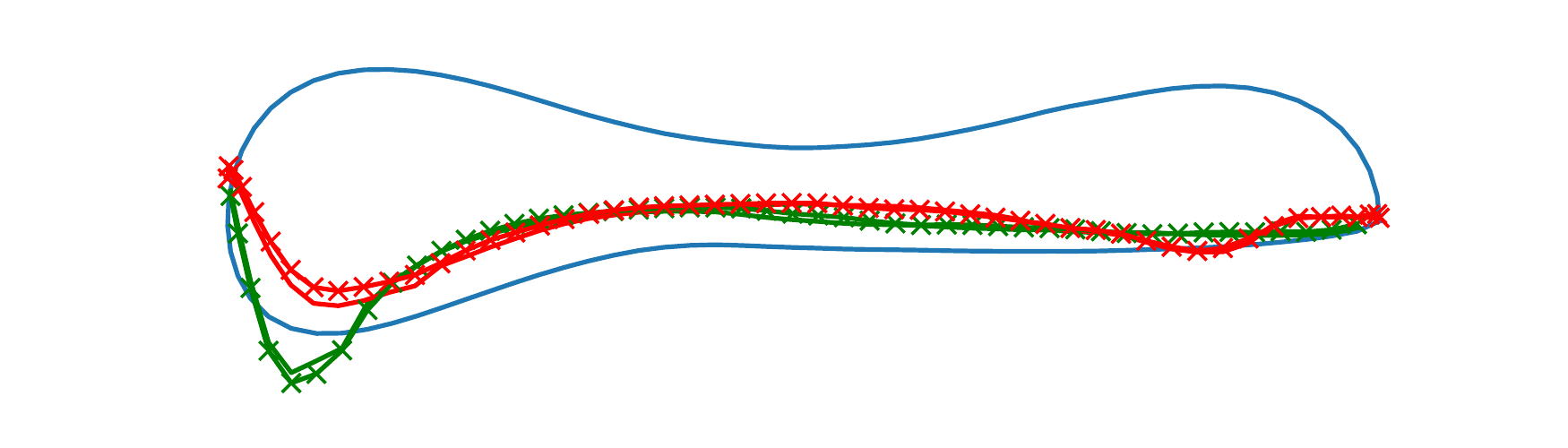}
	\put(3,16){\footnotesize{$F_z = 3.223$, $Z= 3.237$ }}
	\end{overpic}&
\begin{overpic} [scale=0.30,unit=1mm,trim={60 -15 60 -15},clip]{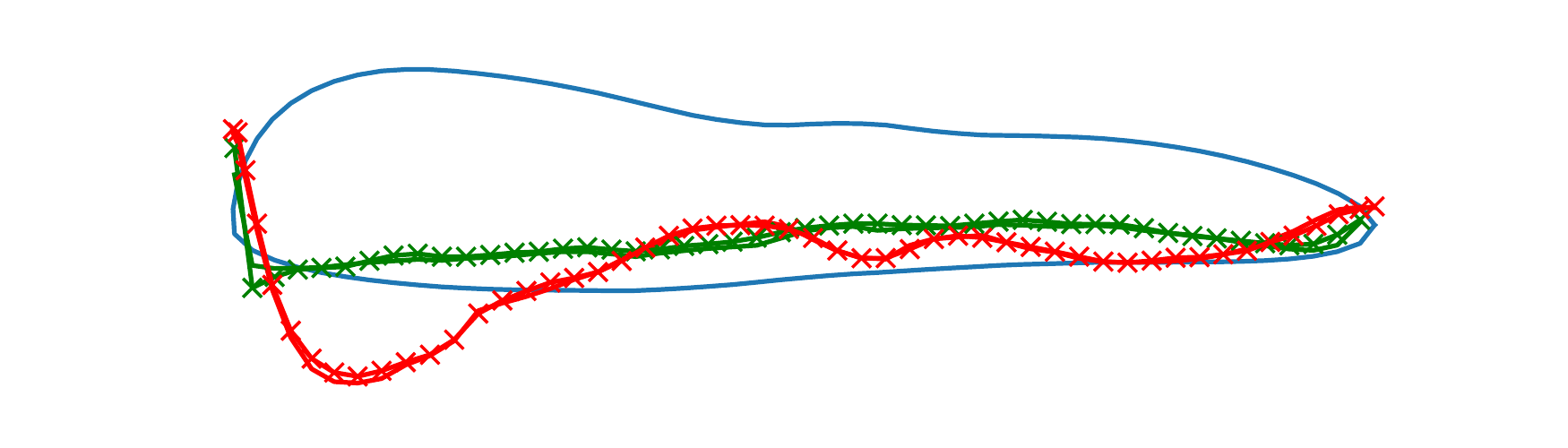}
	\put(3,16){\footnotesize{$F_z = 0.719$, $Z= 0.710$ }}
\end{overpic}\\
\begin{overpic} [scale=0.30,unit=1mm,trim={60 -15 60 15},clip]{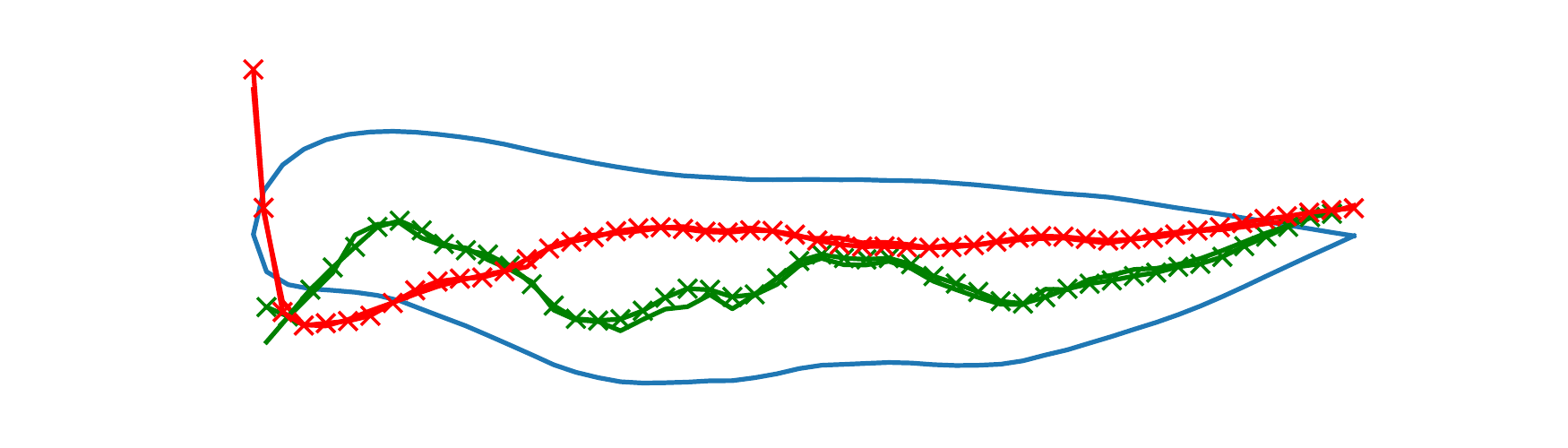}
	\put(3,16){\footnotesize{$F_z = 0.245$, $Z= 0.244$ }}
\end{overpic}&
\begin{overpic} [scale=0.30,unit=1mm,trim={60 -15 60 15},clip]{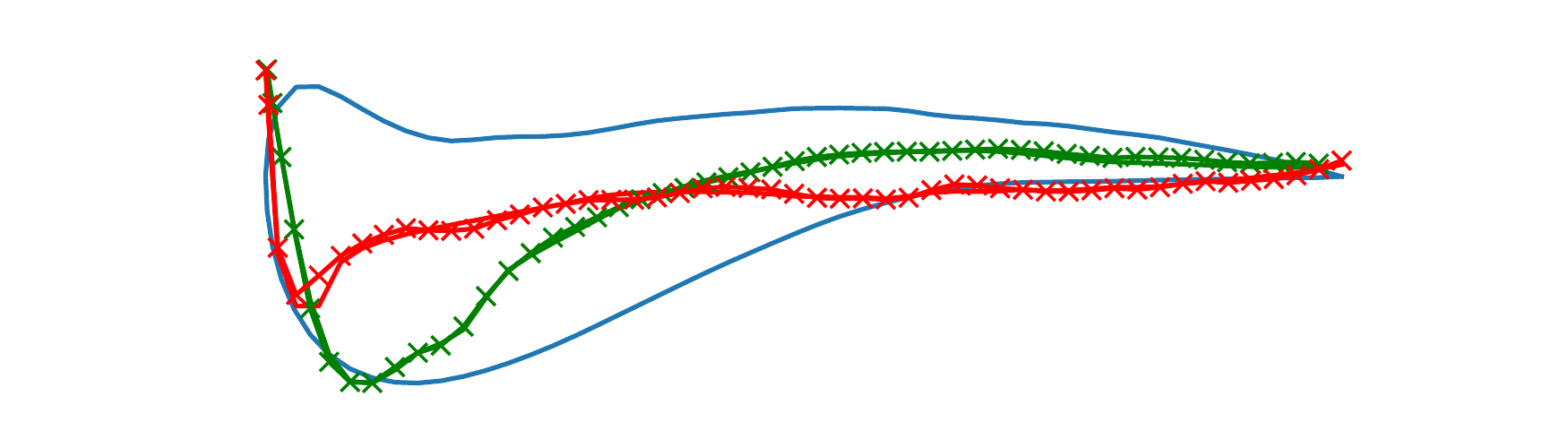}
	\put(3,16){\footnotesize{$F_z = 3.094$, $Z= 3.038$ }}
\end{overpic}
\end{tabular}\
\begin{tabular}{c}
\hline\\
\begin{overpic} [scale=0.25,unit=1mm,trim={50 230 30 320},clip]{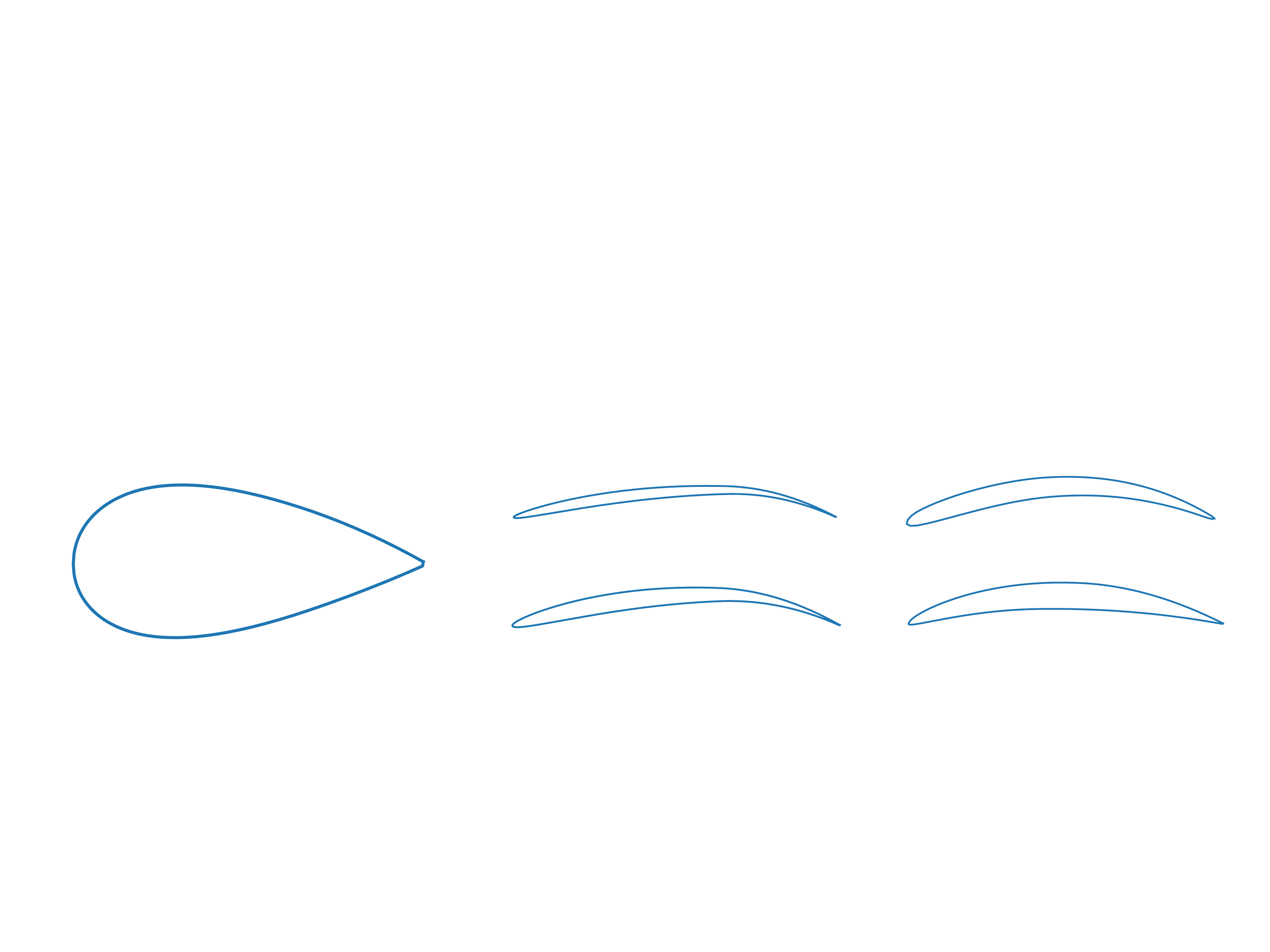}
	\put(1,15){\tiny{Initialization, $\mathcal G = 98.47$ }}
	\put(32,15){\tiny{Ours-NACA, $\mathcal G = -1.23$ }}
	\put(33,7){\tiny{GP-NACA, $\mathcal G = -1.15$ }}
	\put(60,15){\tiny{Ours-18DOF, $\mathcal G = -1.26$ }}
	\put(61,7){\tiny{GP-18DOF, $\mathcal G = -1.10$ }}
\end{overpic}\\
\end{tabular}
\end{center}
        \vspace{-6mm}
	\caption{2D Profiles. (Top) Pressure and drag estimates on four profiles from the testing set, shown in blue. The red and green solid lines depict the ground-truth pressure values above and below. $Z$ is the ground-truth drag. The red and green crosses depict the predicted pressure values, which mostly fall on the corresponding curves. $F_z$ is the predicted drag, which is also close to $Z$. (Bottom) Starting from the shape on the left, we obtain the shapes in the middle using the standard NACA 3-parameter deformation model and the shapes on the right using the more complex 18-parameter model. In both cases, we obtain a lower value of the objective function $\mathcal G$ using our regressor than the GP ones.}
	\vspace{-1mm}
	\label{fig:regressor2D}
\end{figure}

\vspace{-1mm}
\paragraph{Training and testing data.}

As discussed above, airfoil profiles have long been parameterized using three NACA parameters~\cite{Jacobs48}. To generate our training and validation data, we create 8000 training and 8000 testing shapes, such as those depicted by  the blue outlines at the top of Fig.~\ref{fig:regressor2D}. To this end, we randomly select NACA parameters and then further randomize the shape. This is intended to demonstrate that our approach remains effective even when the training shapes belong to a much larger set of shapes that can be far from desirable.  We use the popular CFD simulator \texttt{XFoil} to compute their aerodynamic properties. It takes as input a discretized outline, solves the flow problem using an inviscid-vorticity panel method,  and applies a compressibility correction~\cite{Drela89}. We will demonstrate below that our regressor learned from such non-aerodynamic shapes can nevertheless be used to refine a profile and obtain truly efficient ones, such as those shown at the bottom of Fig.~\ref{fig:regressor2D}.

\vspace{-3mm}
\paragraph{GCNN design choices.}

\begin{figure}[t!]
\begin{center}
	\begin{tabular}{ | l | l | l |}
		\hline
		Model             & $\mathcal A_{C_p}$ & $\mathcal A_{C_D}$ \\ \hline
		\texttt{Standard Separate} & 0.6877             & 0.8223             \\ \hline
		\texttt{Dilated   Separate}   & 0.7442             & 0.8406             \\ \hline
		\texttt{Joint 2 Branches}    & 0.8132             & 0.8490             \\ \hline
		\texttt{Joint 4 Branches} & \textbf{0.8203}    & \textbf{0.8601}    \\ \hline
	\end{tabular}
	\caption{Prediction accuracy for the pressure profile along the airfoil $C_p \in \mathbb R^N$ and the drag coefficient $C_D \in \mathbb R$ on $8000$ randomly generated airfoil shapes. }
	\comment{	\caption{Prediction accuracy for the pressure profile along the airfoil $C_p \in \mathbb R^N$ and the drag coefficient $C_D \in \mathbb R$ on $8000$ randomly generated airfoil shapes. To make the task challenging enough, we generate shapes with a sophisticated random shape generators which draws hundreds of random variables for each new shape and is therefore vastly more complicated than the limited NACA space.}}
\label{tab:prediction}
\end{center}
\end{figure}

We tested several architectures to implement the regressor $F_{\omega}$ of Eq.~\ref{eq:regressor}. 
\vspace{-3mm}
\begin{itemize}
\setlength\itemsep{0.2em}
\item \texttt{Standard Separate}: We use two separate GCNN architectures for drag and pressure prediction. They are exactly the same as discussed in Section~\ref{sec:archi}, except that only one of the final branches is created for each and we use dense $3\times3$ convolutional filters instead of dilated ones.

\item \texttt{Dilated Separate}: We replace the usual convolutions by  {\it{dilated}} ones, which include a spacing between kernel values~\cite{Yu16b}. 

\item \texttt{Joint 2 Branches}: We replace the two separate networks by a shared common branch $\mF_0$ followed by separate branches $F^y$ and $F^z$ for drag and pressure, as discussed in Section~\ref{sec:archi}. 

\item \texttt{Joint 4 Branches}: We push the idea of using separate branches connected to a shared one a step further by adding two more branches that predict the skin friction coefficient along the airfoil and edge fluid velocity. Although these quantities are {\it not} used to compute the objective function, the hope is that forcing the network to predict them helps the joint branch to learn the right features. This is known as disentangling in the Computer Vision literature and has been observed to boost performance~\cite{Caruana97,Rifai12,Ramsundar15}. 

\end{itemize}
We report the accuracy results for these four architectures in Table~\ref{tab:prediction}. As observed by~\citet{Chen15b} for dense semantic image segmentation, the dilated convolutions perform better than the standard ones for regression of dense outputs. Both joint architectures do better than the separate ones, with disentangling providing a further performance boost. At the top of Fig.~\ref{fig:regressor2D}, we superpose the pressure vector computed using the simulator and those predicted by the \texttt{Joint 4 Branches} architecture for 4 different profiles.  

\vspace{-3mm}
\paragraph{Comparing to standard regressors.}

Since our  \texttt{Joint 4 Branches}  GCNN architecture performs best, we will refer to it as \texttt{Ours} and we now compare its accuracy to that of two standard regressors, one based on Gaussian Processes (GPs) and the other on K-Nearest Neighbours (KNNs). For GPs, we use \textit{squared exponential kernels}  because they have recently be shown to be effective for aerodynamic prediction tasks \cite{Toal11, Rosenbaum13,Chiplunkar17}. For KNN regression, we empirically determined that $K = 8$ combined to a distance-based neighbor weighting yielded the best results. Note that in order to compare to such parametric methods, in this experiment only, we restrict our training and test set to the NACA parameter space.

\begin{figure}[t!]
	\centering
	\begin{overpic}[width=80mm, trim={10 20 30 30},clip]{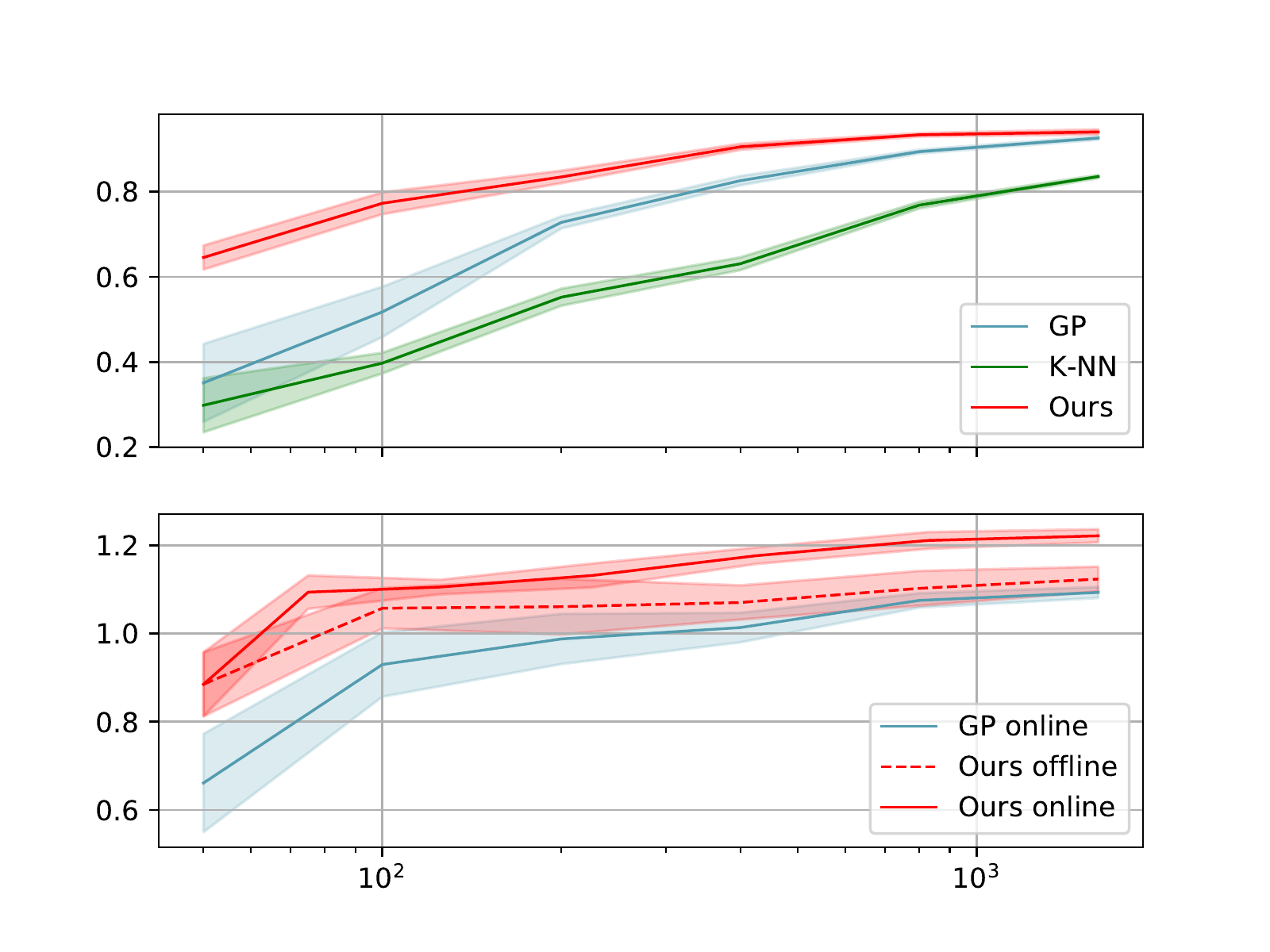}
	\put(-5,40){\footnotesize{$- \mathcal{G}$}}
	\put(-7,120){\footnotesize{$\mathcal{A}_{C_D}$}}
	\put(95,-1){\footnotesize{training set size}}
    \end{overpic}
        \vspace{-3mm}
	\caption{Comparative results for 2D airfoils. (Top) Accuracy of drag prediction.  (Bottom) Shape optimization.} 
        \vspace{-2mm}
	\label{fig:comparison2D}
\end{figure}

The three curves at the top of Fig.~\ref{fig:comparison2D} represent the mean accuracy of the predicted drag and its variance as function of the number of samples used to train the regressors.  Our approach consistently outperforms the other two, especially when there are few training samples. One possible interpretation is that, because our regressor 
operates directly on the shape of the object unlike the other two regress from the NACA parameters, it  learns the local physical interactions between discretization vertices and can therefore  generalize well to unseen shapes.


\vspace{-3mm}
\paragraph{Shape optimization.}

We now use our regressor along with the baseline ones to maximize lift while keeping drag constant. To this end, we take the fitness function of Eq.~\ref{eq:optim} to be
$
\mG (\bY, Z) = - C_L(\bY) + \lambda(Z - Z_{target})^2
$,
where $C_L$ is the function that integrates the pressure values $\bY$ to estimate the lift, $Z_{target}$ is the drag target, and $\lambda$ is a parameter that controls the relative importance of the two terms. In our experiments, we set $\lambda = 100 $ and $Z_{target} = 0.8$.

In the bottom graph of Fig.~\ref{fig:comparison2D}, we show the resulting lift values after performing shape optimization, again as a function of the number of training samples used to train the regressor. The resulting wing profiles are shown at the bottom middle of Fig.~\ref{fig:regressor2D}. 

We also plot the corresponding results of a standard GP-based method, also known as kriging~\cite{Jeong05,Toal11,Nardari17,Xu17}, an industry standard as discussed in Section~\ref{sec:shapeOpt}. Kriging can also be used either offline or online, that is without or with retraining the regressor during the optimization process. To implement the retraining, we relied on an optimal trade-off strategy between exploitation and exploration~\cite{Toal11}. Note that our regressor is good enough to outperform GP online even without retraining. 

In a last experiment, we reparameterized the wing shape in terms of 19 parameters instead of the usual 3 from NACA, as described in detail in the supplementary material. We performed the computation again using this new parameter set in conjunction with either our approach or GPs. The results are shown at the bottom right of Fig.~\ref{fig:regressor2D} and demonstrate our approach's ability to deal with larger models.

\subsection{3D Shapes - Cars}
\label{sec:cars}

\vspace{-1mm}
\paragraph{Training and Testing Data.}

We use the four following datasets
\vspace{-2mm}
\begin{itemize}
\setlength\itemsep{0.2em}
\item \texttt{SYNT-TRAIN} : It is a dataset of 2000 randomly generated 3D shapes such as the one shown at the top of Fig.~\ref{fig:teaser}, which does not need to be  car-like. 

\item \texttt{SYNT-TEST}:  50  more random shapes generated in the same way as those of~\texttt{SYNT-TRAIN} for testing purposes.

\item \texttt{CARS-FineTune} : We downloaded 6 cars CAD models from the web. Two of them are kept for fine-tuning. We augment each model with 3 scaling factors and 9 rotations, to obtain a total of $54$ cars.

\item \texttt{CARS-TEST} : The four remaining CAD models held out for final testing, yielding a total of $108$ shapes after augmentation.

\end{itemize}
\vspace{-2mm}
To generate our random shapes, we introduce a function $f_\bC : \bR^3 \rightarrow \bR^3$, where $\bC$ represents the parameters that control its behavior and apply it to an initially spherical set of vertices $\bX^0$. $f$ is an algebraic function that applies rotations, translations, affine transformations, and dilatations with respect to the center of the shape, which lets us create a wide variety of shapes. The shape at the top Fig.~\ref{fig:teaser} is one of them and we provide more in the supplementary material. We give the precise definition of $f$ in the supplementary material and take  $\bC$ to be a 21D vector. 
We used the industry standard  \texttt{Ansys Fluent}~\cite{Ansys11} to compute their aerodynamic properties with the k-epsilon turbulence model.

\vspace{-3mm}
\paragraph{Comparing to Standard Regressors.}


\begin{figure}[ht!]
\begin{center}
\begin{tabular}{|c|c|c|c|c|c|c|}
\hline
\textbf{}       & {\texttt{VALID}}   &\texttt{TEST}  & {\texttt{TEST-FT}} \\ \hline
\textbf{Method} & $\mathcal A _{C_D} $& $\mathcal A _{C_D}$ & $\mathcal A _{C_D}$ \\ \hline
\textbf{CNN}   &     70.1 \%            &         38.4\%            &         58.1\%              \\ \hline
\textbf{Ours}  & \textbf{77.2\%}     &    \textbf{51.5 \%  }     &    \textbf{70.3\%}   \\ \hline
\end{tabular}
%
%
\end{center}
 \vspace{-0.3cm}
   \caption{Regression results on our three 3D test sets. }
\label{tab:eval:regression3d}
\vspace{-0.3cm}
\end{figure}

In Fig.~\ref{tab:eval:regression3d}, we report the accuracy of our regressor under three different training and testing scenarios: 
\vspace{-2mm}
\begin{itemize}
\setlength\itemsep{0.2em}
\item \texttt{VALID} : The regressor is first trained on \texttt{SYNT-TRAIN} and tested on \texttt{SYNT-TEST}. This is a sanity check since testing is carried out on shapes that have the same statistical distribution as the training ones. 

\item \texttt{TEST} : The regressor is trained on \texttt{SYNT-TRAIN} and tested on \texttt{CARS-TEST}. This is much more challenging since the testing shapes are those of real cars while the training ones are not.

\item \texttt{TEST-FT} : The regressor is trained on \texttt{SYNT-TRAIN}, fine tuned using \texttt{CARS-FineTune}, and tested on \texttt{CARS-TEST}. This is similar to \texttt{TEST} but we help the regressor by giving it a few real car shapes making a few additional epochs of training. 

\end{itemize}
\vspace{-2mm}
Unsurprisingly, the accuracy on  \texttt{TEST} is lower than on  \texttt{VALID}. Nevertheless, fine-tuning with a few car-like examples brings it back up. To assess the importance of using GCNNs instead of regular CNNs, we re-ran all three scenarios using a standard CNN of similar complexity. In other terms, we keep the same architecture where the geodesic convolutions of Eqs.~\ref{eq:interpolate} and~\ref{eq:conv}, are replaced by standard ones. As can be seen on the top row of the figure, the accuracy numbers are much worse in all three cases. 


\begin{figure}[ht!]
	\vspace{-3mm}
	\centering
	\begin{tabular}{c}
	{\includegraphics[width=60mm, trim={0 0 0 0},clip]{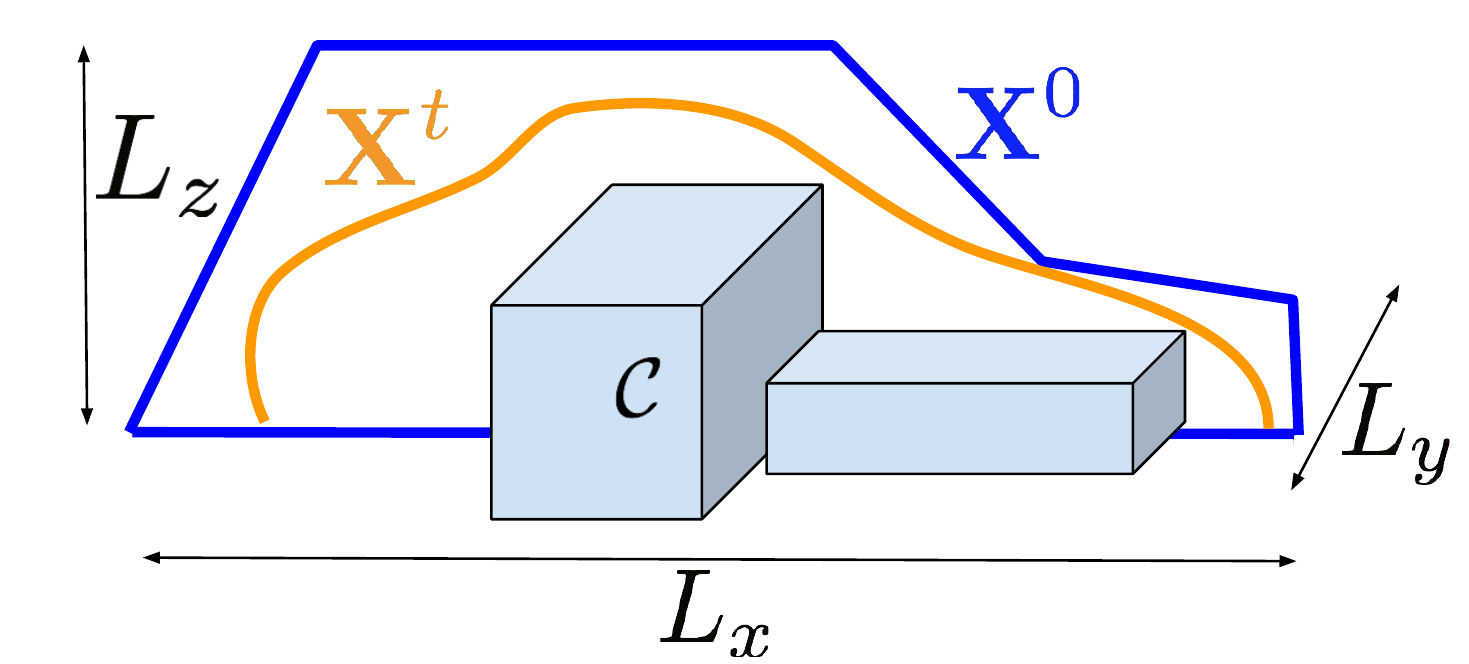}}\\
	\begin{tabular}{|c|c|c|c|}
        \hline
        \textbf{Method} & Init & 50 Sim  & 100 Sim  \\ \hline
        \textbf{Ours - Online}  & 80.2 &\textbf{4.91} & \textbf{ 4.71}      \\ \hline
        \textbf{Ours - OffLine}  & 80.2 & 7.56 & 7.56      \\ \hline
        \textbf{GP - Online}  &  80.2 & 16.65 &  12.56      \\ \hline
        \end{tabular}
	\end{tabular}
	\vspace{-2mm}
	\caption{3D car shape optimization. (Top) Feasible shapes must remain out of the two parallelepipeds. (Bottom) Drag of the optimized shapes given the number of  calls to the simulator for the online methods. }
	\vspace{-3mm}
	\label{fig:car}
\end{figure}

\vspace{-3mm}
\paragraph{Shape Optimization}

In this section, we use the GCNN regressor  pre-trained on our \texttt{SYNT-TRAIN} data to minimize the drag generated by a car-like shape, that is,
$
\mG (\bY, Z) =  Z
$.
Without constraints, the shape would collapse to an infinitely thin one. To allows for passengers and an engine, we need the constraint $\mC$ of Eq.~\ref{eq:optim}.  Feasible shapes are defined as those that remain outside of two parallelepiped, one for the engine and the other for the passenger compartment, as shown at the top of Fig.~\ref{fig:car}. 

As in the case of the airfoils, the GP regressor takes as input the 21 parameters $\bC$  of the deformation function $f_\bC$ introduced above while ours operates directly on the surface mesh. The initial shape $\bX^0$ that $f_\bC$ operates on is the CAD model of a car shown on the left at bottom of Fig.~\ref{fig:teaser} and the result is shown on the right. At the bottom of Fig.~\ref{fig:car}, we report the resulting drag for a given number of call to the simulator during the minimization, 50 or 100 for the online methods and 0 for the offline one. 

Again whether offline or online, our approach outperforms the online version of GP and finds a better optimum for the $21$ parameters. In the online case, note that with {\it only} 50 calls to the simulator our result is already close to the optimum even though our network, pre-trained on \texttt{SYNT-TRAIN}, had {\it never} seen a  car before. 


\section{Conclusion}
\blfootnote{ This work was realized with the help of the software ANSYS by ANSYS Inc.}

We have shown that we could first train Geodesic CNNs to reliably emulate a output of a CFD simulator and then use them to optimize them the aerodynamic performance of a shape. As a result, we can outperform state-of-the-art techniques by 5 to 20\% on relatively simple 2D problems and solve previously unsolvable ones. 

In our current implementation, we use parameterized models to reduce the shape's number of degrees of freedom and make the optimization problem tractable. Since our method can operate on generic meshes, in future work, we intend to use the  Laplacian parameterization introduced by~\citet{Ngo16} to increase or decrease the number of degrees of freedom at will and make our approach fully flexible. 

\label{sec:parametrization}

\bibliography{string,vision,learning,optim,cfd,graphics}
\bibliographystyle{icml2018}
\section{Appendix}
In this supplementary material, we first provide additional detail on the handling of the irregular vertices of the Cube-Mesh CNNs of Section 3.1. We then give analytical definitions of the 2D and 3D deformation parameterizations of Sections 5.1 and 5.2. 

\subsection{Handling Singular Points for Semi-Regular Quad-Meshes}

As discussed in Section 3.1 of the paper,  when mapping a surface onto a cube-mesh, we have to deal with irregular vertices, which correspond to the corners of the cube and  have three neighbors instead of four. To perform convolutions efficiently we first unfold the cube surface onto a plane. As illustrated by Fig.~\ref{fig:singular}, we can then simply pad irregular corners with the feature values associated to cube edges. This enables us to use standard convolutional kernels even in the neighborhood of irregular vertices. Furthermore, since we use Geodesic Convolutions, the irregularity is naturally handled by the interpolation operation.

\begin{figure}[ht!]
	\centering
		{\includegraphics[width=80mm, trim={0 0 0 0},clip]{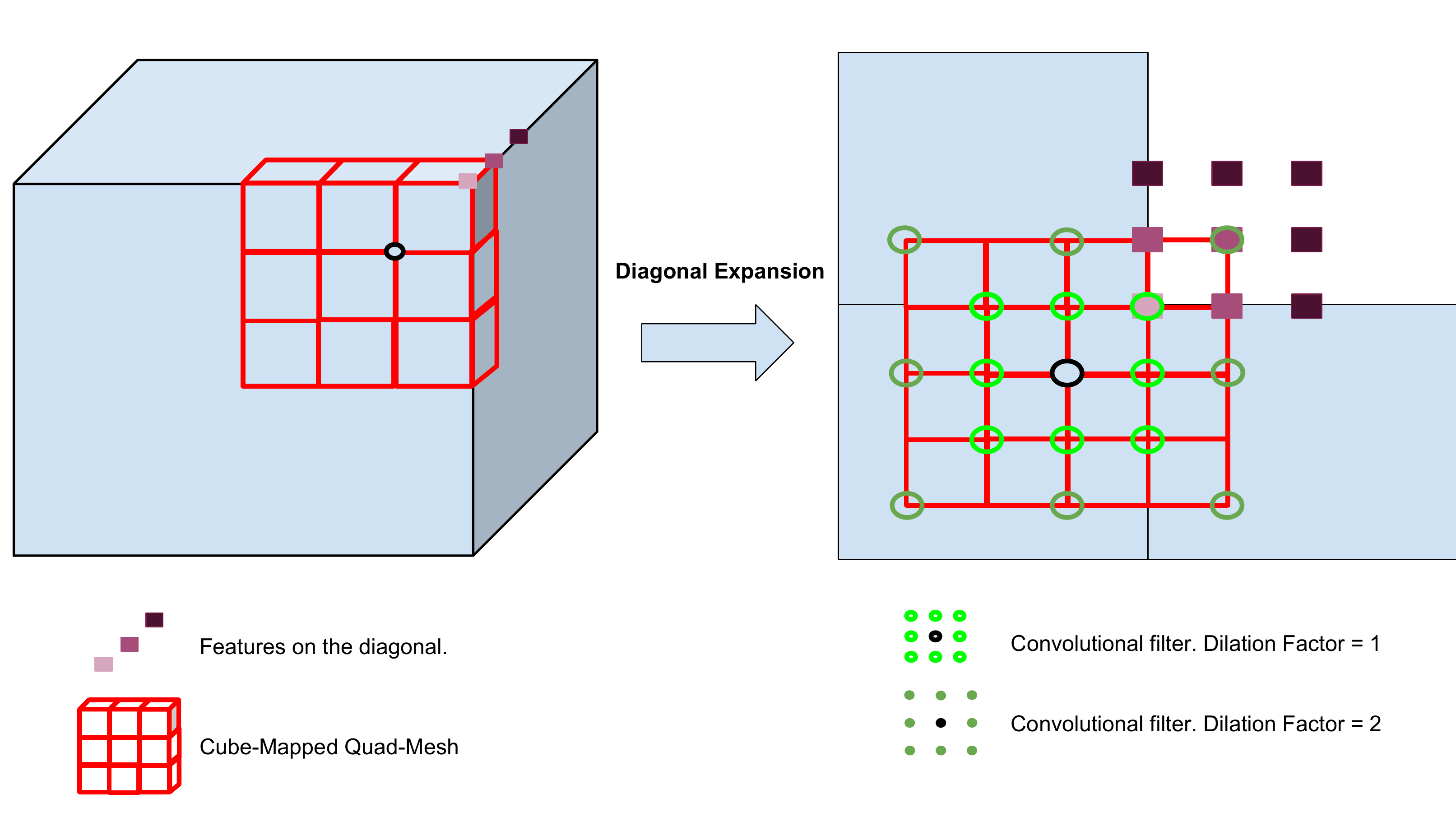}}
	\vspace{-4mm}
	\caption{Handling the singularities of the Quad-Mesh for convolution purposes.}
	\label{fig:singular}
\end{figure}

\comment{
\begin{figure*}[ht!]
	\centering
	\begin{overpic}[width=160mm,clip]{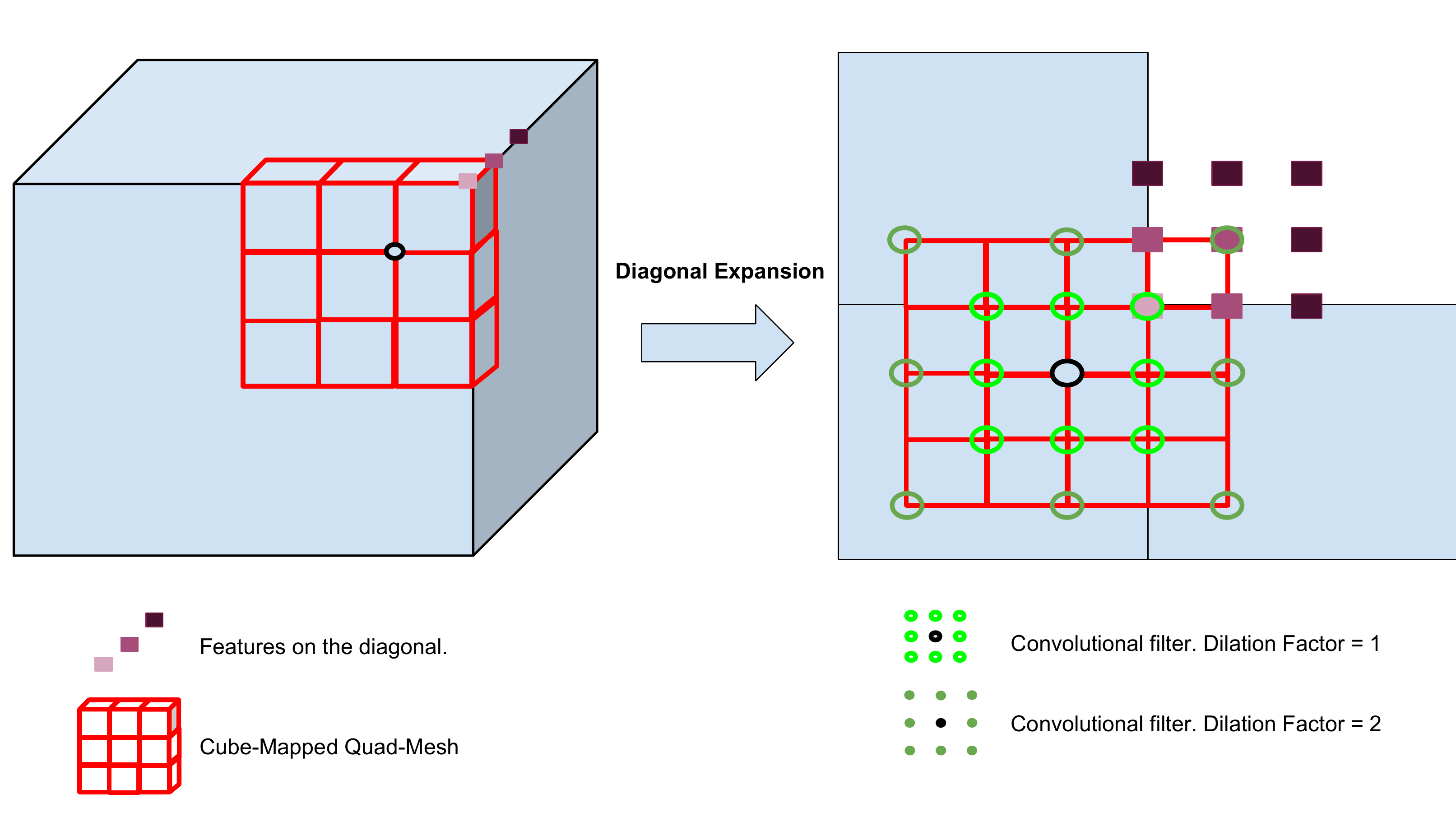}
	\end{overpic}
	\caption{Description of our handling of the singularities of the Quad-Mesh for the convolutions.}
	\label{fig:singular}
\end{figure*}
}

\subsection{Airfoil Parameterization in 2D}

In this section we will first briefly describe the standard NACA airfoil 4 digit parameterization~\cite{Jacobs48}, which, confusingly involves 3 degrees of freedom.  We then discussed our extension to 19 degrees of freedom. 

\paragraph{NACA 4 digit.}

Without loss of generality, we can assume that the airfoil is of unitary cord length and let  $0 \leq x \leq 1$ the coordinate that defines the position along that length. Let us further consider the airfoil thickness $t$, maximum camber $m$ , along with its location $p$. To compute the airfoil shape, we first define the mean camber line 
\begin{align*}
y_c &= \left\{ \begin{array}{ll}
\displaystyle{\frac{ m}{ p     ^2} \left( 2p x- x^2   \right)}, & 0 \leq x \leq p \\
\\
\displaystyle{\frac{ m}{(1 - p)^2} \left( (1 - 2p) + 2p x - x^2 \right)}, & p \leq x \leq 1
\end{array} \right.\\
\end{align*}
%
and the airfoil thickness to camber $y_t$ as
\begin{small}
\[
5t \left[ 0.2969 \sqrt{x} -0.1260 x -0.3516 x^2 + 0.2843 x^3 -0.1015 x^4 \right] .
\]
\end{small}
Since the thickness needs to be applied perpendicular to the camber line, the coordinates $(x_U,y_U)$ and $(x_L,y_L)$, of the upper and lower airfoil surface, respectively, become
\begin{align}
x_U &= x   - y_t\, \sin \theta, \qquad &
y_U &= y_c + y_t\, \cos \theta, \\
x_L &= x   + y_t\, \sin \theta, &
y_L &= y_c - y_t\, \cos \theta,
\end{align}
where
\begin{align}		
\theta &= \arctan{ \left( \frac{dy_c}{dx} \right)},\\
\frac{dy_c}{dx} &= \left\{\begin{array}{ll}
			\displaystyle{\frac{2m}{p^2} \left(p - x\right)}, & 0 \leq x \leq p \\
			\\
			\displaystyle{\frac{2m}{(1 - p)^2} \left(p - x\right)}, & p \leq x \leq 1
		\end{array} \right. 
\end{align}
Thus, the wing shape is entirely defined by the choice of $t$, $m$ , and $p$.

\paragraph{18-parameter foils.}

We increase the number of degrees of freedom by that writing the 3 parameters $t$, $m$, $p$ as quadratic functions of $x$, that is, 
\begin{align*}
t(x) &= t_0 + t_1 x + t_2 x^2 \; \\
m(x) &= m_0 + m_1 x + m_2 x^2  \; \\
p(x) &= p_0 + p_1 x + p_2 x^2  \; \\
\end{align*}
where the the $p_i$, $m_i$, and $q_i$ control the new degrees of freedom. Moreover we allow the lower and upper airfoil surfaces to be associated two two different camber lines, hence doubling the total number of degrees of freedom to $2 \times (3+3+3)$.

\subsection{Surface Parameterization in 3D}

As discussed in Section 5.2, we parametrize 3D shape deformations using a  transformation function $f_{\bC} : \bR^3 \rightarrow \bR^3$ that applies to the vertices of an initial shape $\bX^0$, where $\bC$ is a 21D vector.  For clarity, let us split the 21 components of $\bC$ into three groups, one for each axis $\bC = \{C^x_i\}_{i = 0 \dots 6} \cup \{C^y_i\}_{i = 0 \dots 6} \cup \{C^z_i\}_{i = 0 \dots 6} $. As show in Fig.~\ref{fig:car}, $L_x,L_y,L_z$, denote the maximal size over each dimension and let $(x,y,z)$ be the coordinates of a specific vertex $X$. We write
\begin{align*}
f_\bC(X)_x &= C^x_0 + x [ C^x_1 + C^x_2x \\
& + C^x_3cos(\dfrac{y}{L_y} 2 \pi) + C^x_4cos(\dfrac{z}{L_z} 2 \pi)  \\
& + C^x_5sin(\dfrac{y}{L_y} 2 \pi) + C^x_6sin(\dfrac{z}{L_z} 2 \pi) ] \;, \\
f_\bC(X)_y &= C^y_0 + y[C^x_1 + C^y_2y \\
& + C^y_3cos(\dfrac{x}{L_x}  \pi) + C^y_4cos(\dfrac{x}{L_x} 2 \pi)  \\
& + C^y_5sin(\dfrac{x}{L_x}  \pi) + C^y_6sin(\dfrac{x}{L_x} 2 \pi)]\;,\\
f_\bC(X)_z &= C^z_0 + z[C^x_1 + C^z_2z \\
& + C^z_3cos(\dfrac{x}{L_x}  \pi) + C^z_4cos(\dfrac{x}{L_x} 2 \pi)  \\
& + C^z_5sin(\dfrac{x}{L_x}  \pi) + C^z_6sin(\dfrac{x}{L_x} 2 \pi)]\;.
\end{align*}
This simple parametric transformation provides enough freedom to generate sophisticated shapes. Furthermore, the initial shape corresponds to setting all the parameters to 0, except from $C^x_1,C^y_1,C^z_1$, which are set to 1.

%

%
%
%
%

\end{document}